# Exploring Information Retrieval Landscapes: An Investigation of a Novel Evaluation Techniques and Comparative Document Splitting Methods

**Esmaeil Narimissa and David Raithel**

## Abstract

The performance of Retrieval-Augmented Generation (RAG) systems in information retrieval is significantly influenced by the characteristics of the documents being processed. In this study, the structured nature of textbooks, the conciseness of articles, and the narrative complexity of novels are shown to require distinct retrieval strategies. A comparative evaluation of multiple document-splitting methods reveals that the Recursive Character Splitter outperforms the Token-based Splitter in preserving contextual integrity. A novel evaluation technique is introduced, utilizing an open-source model to generate a comprehensive dataset of question-and-answer pairs, simulating realistic retrieval scenarios to enhance testing efficiency and metric reliability. The evaluation employs weighted scoring metrics, including SequenceMatcher, BLEU, METEOR, and BERT Score, to assess the system's accuracy and relevance. This approach establishes a refined standard for evaluating the precision of RAG systems, with future research focusing on optimizing chunk and overlap sizes to improve retrieval accuracy and efficiency.

# 1. Introduction

The emergence of Retrieval-Augmented Generation (RAG) systems marks a significant development in information retrieval and natural language processing. RAG systems combine retrieval-based and generative machine learning models to improve the contextual relevance of responses to complex queries, drawing from extensive text corpora. Recent research has demonstrated the effectiveness of such systems in various applications, from question-answering tasks to knowledge-based reasoning [1]. However, the efficiency of these systems depends heavily on the structure of the documents being processed and the methods employed for splitting and retrieving relevant text segments [2].

RAG systems are particularly suited for handling unstructured data, such as textbooks, articles, and novels, each presenting distinct retrieval challenges. Prior studies have shown that structured data, like textbooks, allow for easier segmentation, whereas narrative content in novels introduces complexities in maintaining contextual integrity [3]. This research focuses on comparing document-splitting techniques, specifically the Recursive Character Splitter and Token-based Splitter, to assess their performance in preserving text coherence and retrieval accuracy. The Recursive Character Splitter is hypothesized to perform better based on its ability to maintain context across text chunks [4].

The system's operational environment utilizes state-of-the-art technologies, including OpenAI's GPT models for generative tasks and Pinecone's vector database for high-speed text retrieval [5]. Additionally, open-source models from LM Studio are integrated to provide alternative methods for document embedding and retrieval, as explored in recent work on hybrid retrieval architectures [6]. This combination allows for comprehensive experimentation across multiple approaches and facilitates the handling of large text corpora with privacy considerations in mind.

Exploratory data analyses reveal performance variations across document types and splitting methods. The Recursive Character Splitter consistently outperforms the Token-based Splitter by maintaining greater contextual continuity, particularly with complex documents such as novels [3]. Additionally, the comparative evaluation of retrieval methods highlights the strengths and limitations of both OpenAI and LM Studio's embeddings. These findings contribute to ongoing research on optimizing RAG systems for various document types and retrieval tasks [1].

This article aims to assess the impact of document characteristics on the performance of RAG systems, with a focus on comparing document-splitting methods and evaluating retrieval techniques. Section 2 provides an overview of the system architecture and implementation, detailing the use of large language models (LLMs) and vector databases. Section 3 discusses the document analysis framework, including the selection of document types and the splitting techniques employed. Section 4 introduces the evaluation methodology, highlighting the novel approach used to generate question-and-answer pairs for performance assessment. The results and discussion are presented in Section 5, where the impact of splitting methods and retrieval techniques is analyzed across different document types. Finally, Section 6 concludes the study with key findings and suggests future research directions for optimizing retrieval strategies in RAG systems. All implementation details, including the code used for evaluation and data preprocessing, can be found in the GitHub repository.

## 2. System Implementation and Overview

OpenAI's GPT models and Pinecone's vector database are utilized to establish a scalable and efficient foundation for complex text processing within the RAG system. Data ingestion and preprocessing form the core of the operational workflow, with text extracted from PDF documents using the PyPDF2 library. Rigorous preprocessing ensures data cleanliness and uniformity, facilitating effective segmentation and transformation into structured formats suitable for analysis and retrieval.

The system incorporates renowned document-splitting techniques, including Recursive Character Splitters and Token-based Splitters, to enhance semantic completeness and maintain narrative flow. These splitting methods enable the transformation of text chunks into dense vector representations, improving retrieval and comparison efficiency. This approach optimizes interaction with large text datasets by employing advanced techniques in retrieval, embedding, and generation.

The full codebase implementing the RAG system, including the use of OpenAI's GPT models and document splitting techniques, is available at GitHub repository.

## 3. Document Analysis Framework

### 3.1 Document Types Used

A diverse set of document types is utilized to evaluate the performance of the RAG system, ensuring comprehensive analysis. These types include a physics journal review-article [7], a scientific textbook [8], and the novel *War and Peace* by Leo Tolstoy, each presenting distinct challenges for information retrieval. The journal article provides dense scientific content with specialized terminology, while the textbook offers structured technical information, suitable for evaluating the system's handling of educational materials. The novel, with its intricate narrative and complex character interactions, tests the system's ability to manage and retrieve information from narrative-driven content.

The selection of these document types highlights the varying performance of the RAG system. Textbooks and scientific articles typically yield higher retrieval scores due to their structured format and dense information, which align well with the system's capabilities in processing complex terminologies. In contrast, novels present challenges due to their expansive and less defined segments, resulting in lower performance scores and necessitating adaptations to handle narrative content effectively. This variation underscores the importance of designing retrieval systems versatile enough to handle diverse document types, providing critical insights for future system optimizations aimed at improving retrieval accuracy and efficiency.

## 3.2 Document Splitting Techniques and Embedding Generation

Managing the large size of text documents presents a primary challenge in the development of the RAG system. To address this, two document-splitting techniques are employed, each tailored to specific content requirements. A standard chunk size of 1000 characters and an overlap of 200 characters are used to ensure continuity and context preservation.

The Recursive Character Splitter segments text based on a fixed character count, maintaining narrative flow and ensuring that no critical information is lost at chunk boundaries. This method is particularly effective for documents requiring high levels of contextual integrity, such as novels or detailed analytical pieces. The 200-character overlap further supports narrative and argumentative coherence.

In contrast, the Token-based Splitter focuses on semantic completeness, dividing documents based on token patterns rather than size. This approach is suited for handling complex structures within technical documents or research papers, where preserving semantic integrity is crucial. The overlap ensures that tokens at the edges of chunks retain their contextual meaning, essential for accurate retrieval.

Following the splitting process, each text chunk is transformed into dense vector representations using embedding models. These embeddings capture the semantic essence of the text, enhancing the system's capacity for detailed retrieval and comparison. By converting the text into machine-interpretable formats, the system's retrieval operations are made more efficient.

The integration of Pinecone, a vector database optimized for scalability and speed, facilitates rapid retrieval of relevant text chunks based on query embeddings, streamlining the response generation process. This efficiency is crucial in applications requiring both speed and accuracy, such as real-time information retrieval and interactive AI systems. The strategic application of document splitting and embedding generation is central to the RAG system, enabling efficient handling of complex modern data.

## 3.3 Overview of Retrieval Techniques

Two distinct retrieval techniques are employed in the RAG system to optimize performance across various document types and queries. The first method utilizes OpenAI's text embeddings, which encode text into high-dimensional vectors, effectively capturing semantic nuances. This technique enables precise matching of queries with relevant document chunks, ensuring high relevance and accuracy in the responses. Its ability to retrieve contextually similar text chunks based on query embeddings further enhances the system's performance.

The second retrieval approach leverages LM Studio's "bge-large-en-v1.5-gguf" open-source embedding model [9], providing an alternative strategy for text matching and embedding. This method complements OpenAI's embeddings by excelling at handling diverse document types and splitting methods, adding versatility to the retrieval process. The combination of both techniques equips the system to address a wide array of queries, improving flexibility and retrieval effectiveness.

Comparative analysis of the two methods indicates that both are effective, though performance varies depending on document type and splitting method. OpenAI's approach consistently outperforms LM Studio's in certain scenarios due to its superior ability to capture semantic nuances. Understanding these strengths and limitations allows retrieval strategies to be tailored for specific needs, ensuring optimal performance across use cases. The strategic use of multiple retrieval techniques enhances the system's overall effectiveness and provides insights for future developments, including the integration of adaptive techniques and hybrid models.

# 4. A novel Methodology for System Evaluation

## 4.1 Automated Q&A Generation for RAG Performance Assessment

An automated system was implemented to generate a substantial dataset of question-and-answer pairs, essential for evaluating the performance of various RAG settings. LM Studio's open-source large language model, "TheBloke/dolphin-2.6-mistral-7B-GGUF," [10] was utilized to systematically generate questions and answers from selected chunks of textbooks, articles, and novels. This process mimics realistic information retrieval scenarios, allowing for rigorous assessment of the accuracy and effectiveness of the document processing strategies employed within the RAG system.

The approach dynamically generates content from a diverse range of textual materials, creating a broad testing environment. Sequential operations were conducted, starting with the identification of optimal text segments for question generation, followed by the model producing relevant questions and answers. The results were compiled into an Excel dataset, "eval_ds.xlsx," which served as a crucial resource for subsequent performance analyses. This automation enhances testing efficiency and ensures that performance metrics are based on consistently generated, controlled data, significantly improving the reliability of the findings.

## 4.2 Optimized Scoring Techniques for RAG Systems Evaluation

A novel multi-metric scoring approach has been adopted to evaluate the outputs of the RAG system, designed to assess the quality of responses across dimensions of accuracy and relevance. The evaluation framework incorporates four primary metrics: SequenceMatcher [11], BiLingual Evaluation Understudy (BLEU) [12], Metric for Evaluation of Translation with Explicit ORdering (METEOR) [13], and Bidirectional Encoder Representations from Transformers (BERT) Score [14]. Each metric evaluates a distinct aspect of text quality. SequenceMatcher measures exact string matches, providing a means to verify literal accuracy and the presence of specific keywords. BLEU assesses n-gram overlap, offering insights into fluency and syntactical correctness, while METEOR considers synonyms and stemming, aligning closely with human judgment on content relevance and meaning.

The integration of the BERT Score introduces a significant advancement, allowing for the analysis of contextual relationships between words in generated and reference texts. BERT Score is particularly adept at evaluating the semantic depth of responses by using bidirectional representations to provide a nuanced view of contextual accuracy. This metric complements the others by adding a crucial layer of semantic evaluation, essential for understanding the RAG system's effectiveness in generating contextually appropriate answers.

Each metric has been strategically weighted to align with the specific requirements of the RAG system. SequenceMatcher and BLEU, with weights of 0.30 each, prioritize exact matches and literal accuracy, critical for precise information retrieval. METEOR, assigned a weight of 0.20, balances form and meaning, while BERT Score, also weighted at 0.20, focuses on semantic similarity. Though important, semantic similarity is considered less critical than factual accuracy in this context.

This novel weighted scoring approach enhances the system's ability to discern the quality and relevance of outputs, providing detailed feedback that allows for continuous system improvements. By optimizing these metrics to suit the characteristics of the RAG system, errors are minimized, ensuring a high standard of response accuracy. This evaluation framework is essential for the practical application of retrieval-augmented technologies in complex information environments.

# 5. Results & Discussions

## 5.1 Overview of Applied Exploratory Data Analysis (EDA)

EDA techniques were crucial in assessing the effectiveness and refinement of the RAG system. Several analytical methods, including Descriptive Statistics, Analysis of Variance (ANOVA) [15], and pairwise comparisons, were employed to examine performance metrics derived from various scoring techniques. These methods provided insights into the dataset, which comprised document types, splitting methods, and retrieval scores from the LM Studio and OpenAI systems.

Descriptive Statistics offered an initial overview of the data distribution across document types (Novels, Textbooks, and Articles) and splitting methods. This analysis revealed key characteristics of the dataset, such as the distribution of average scores for each retrieval method, highlighting variability based on document type and other factors. This preliminary analysis identified areas requiring deeper investigation, setting the foundation for more complex statistical tests.

ANOVA was instrumental in determining whether differences in retrieval scores across document types and between retrieval methods were statistically significant. The results indicated significant variations in the performance of the LM Studio and OpenAI systems, confirming that document type affects retrieval efficacy. These insights are critical for optimizing retrieval algorithms by identifying document types that require specific adjustments.

Pairwise comparisons further explored the differences between document types within each retrieval method. This analysis provided clarity on which document types consistently yielded higher or lower scores, offering a clear understanding of the relative strengths and weaknesses of each retrieval system. These findings are essential for refining the RAG system by revealing how content structure and complexity influence retrieval success.

In summary, the EDA techniques each contributed uniquely to understanding the RAG system's performance across varied datasets. Descriptive Statistics provided a broad overview, ANOVA identified significant trends, and pairwise comparisons detailed specific document-type variations.

## 5.2 Descriptive Statistics

Descriptive statistics, as part of the EDA, offer insights into the distribution of final average scores across the two retrieval methods, LM Studio (LMS) and OpenAI (OAI). The histograms in **Figure 1** illustrate the distribution patterns for both methods, providing a visual representation of the performance differences.

The document type distribution within the dataset is relatively balanced, with 100 entries for novels, 98 for textbooks, and 97 for articles. This even distribution across the three content types allows for a detailed analysis of performance metrics across different document genres. Such balance establishes a solid foundation for a comparative analysis of retrieval performance between the document types.

Regarding the splitting methods, the Recursive Character Splitter (RCS) and the Token Text Splitter (TTS) are applied almost evenly across each document type. This balanced approach enables a focused examination of the impact of each splitting technique, facilitating a precise comparison of performance metrics between RCS and TTS.

The distribution of final average scores for LMS reveals a concentration of lower values, with a few high-score outliers. This pattern suggests challenges in LMS's retrieval accuracy, indicating that the method underperforms in certain contexts. In contrast, OAI demonstrates a more effective and consistent performance, with its scores clustering predominantly at higher values. The broader spread and higher peak in OAI's score distribution suggest that it handles variability in document types and complexities more adeptly than LMS.

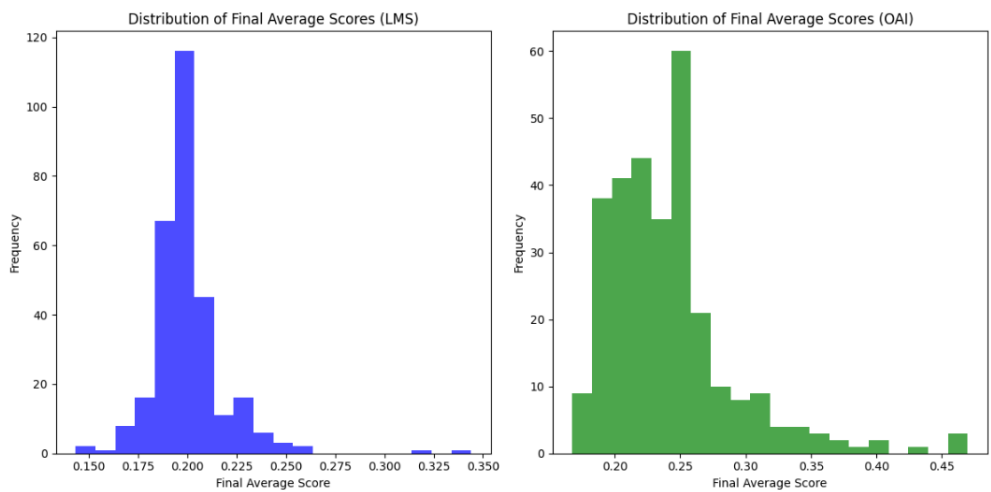

**Figure 1**. Distribution of Final Average Scores for LMS and OAI

## 5.3 Performance Comparison by Document Type

The analysis of average final scores across different document types reveals distinct patterns in the performance of the LMS and OAI retrieval methods. The bar chart in **Figure 2** illustrates how these methods perform across articles, novels, and textbooks, highlighting key differences in retrieval efficiency based on content type. The analysis indicates that articles outperform novels across both retrieval methods, demonstrating the advantages of structured and concise content in enhancing retrieval efficiency. When comparing articles and textbooks, the performance is similar in the LMS method, though articles slightly outperform textbooks with the OAI method, underscoring the effectiveness of OAI in handling concise content. Textbooks, in turn, achieve slightly better scores than novels, further emphasizing the benefits of structured content in both retrieval systems.

The ANOVA results, presented in **Table 1**, statistically validate the performance differences across document types for both LMS and OAI methods. For LMS, the ANOVA test reveals a statistically significant difference in retrieval scores across document types, with a p-value of 0.000005, confirming that document type markedly influences LMS performance. Similarly, the ANOVA test for OAI retrieval scores indicates a significant difference across document types, with a p-value of 0.000001, demonstrating that document type significantly affects OAI's retrieval performance.These statistical results, along with the visual data from the histograms, confirm that both LMS and OAI retrieval methods exhibit significant differences in performance across various document types. The p-values of 0.000005 for LMS and 0.000001 for OAI underscore the crucial role that document type plays in influencing the retrieval efficiency and accuracy of RAG systems.

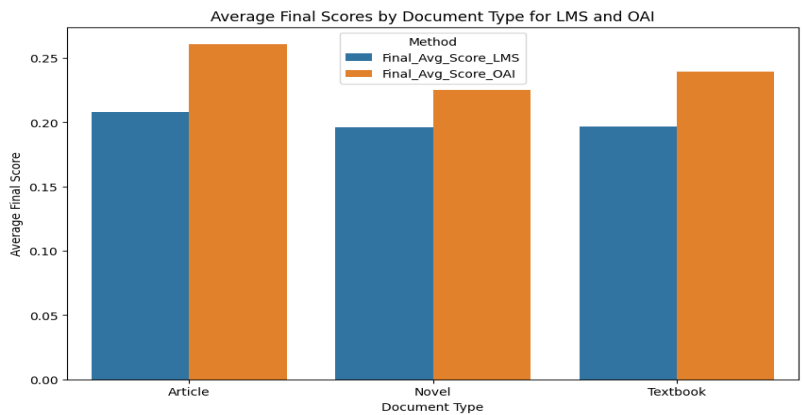

**Figure 2**. Average Final Scores by Document Type for LMS and OAI across various document types.

**Table 1.** ANOVA Results for Retrieval Scores by Document Type

| Source of Variation | Sum of Squares | Degrees of Freedom | F-Value | p-Value |
|---|---|---|---|---|
| LMS Document Type | 0.008582 | 2 | 12.655763 | 0.000005 |
| LMS Residual | 0.099010 | 292 | N/A | N/A |
| OAI Document Type | 0.062681 | 2 | 14.179976 | 0.000001 |
| OAI Residual | 0.645378 | 292 | N/A | N/A |

## 5.4 Pairwise Comparisons between Document Types for Retrieval Scores

The Tukey HSD (Honestly Significant Difference) test [16], a post-hoc analysis, was employed to identify significant differences between means for retrieval scores across document types. **Table 2** presents the results of the Tukey HSD test for both LMS and OAI methods, reporting mean differences, adjusted p-values (p-adj) for multiple comparisons, confidence interval bounds (Lower, Upper), and whether the null hypothesis of equal means was rejected.

The LMS retrieval scores reveal notable differences across document types. Articles consistently outperform novels, with a statistically significant mean difference of -0.0117 and a p-value of less than 0.0001. Similarly, articles also surpass textbooks, with a significant mean difference of -0.0112 and a p-value of 0.0001. However, no significant difference is observed between novels and textbooks, as indicated by a p-value of 0.9807.

For the OAI retrieval scores, significant disparities between document types are also evident. Articles outscore novels with a notable mean difference of -0.0355 and a p-value of less than 0.0001. Articles also perform better than textbooks, with a mean difference of -0.0214 and a p-value of 0.0047. While the difference between novels and textbooks is not statistically significant, it approaches significance, with novels slightly outperforming textbooks by a mean difference of 0.0141 and a p-value of 0.0905.

These findings confirm that document type significantly impacts retrieval performance across both retrieval systems, with articles consistently achieving higher scores. This pattern suggests that the structured and clear content of articles facilitates more effective retrieval compared to the narrative and variable structures found in novels and textbooks. Such insights could inform the optimization of retrieval strategies and algorithmic adjustments tailored to specific document types.

**Table 2.** Pairwise Comparisons of Retrieval Scores by Document Type

LMS Retrieval Scores

| Group1 | Group2 | Mean Difference | p-adj | Lower | Upper | Reject |
|--------|--------|-----------------|-------|-------|-------|--------|
| Article | Novel | -0.0117 | 0.0000 | -0.0179 | -0.0055 | True |
| Article | Textbook | -0.0112 | 0.0001 | -0.0174 | -0.005 | True |
| Novel | Textbook | 0.0005 | 0.9807 | -0.0057 | 0.0067 | False |

OAI Retrieval Scores

| Group1 | Group2 | Mean Difference | p-adj | Lower | Upper | Reject |
|--------|--------|-----------------|-------|-------|-------|--------|
| Article | Novel | -0.0355 | 0.0000 | -0.0512 | -0.0197 | True |
| Article | Textbook | -0.0214 | 0.0047 | -0.0373 | -0.0055 | True |
| Novel | Textbook | 0.0141 | 0.0905 | -0.0017 | 0.0298 | False |

## 5.5 Impact of Splitting Methods

**Table 3** presents the count distribution of document splitting methods alongside the comparative average final scores and ANOVA results for both LMS and OAI retrieval methods, highlighting the statistical significance and performance differences between the RCS and the TTS.

**Figure 3** illustrates the average final scores obtained using the RCS and the TTS for both LMS and OAI retrieval methods, demonstrating the superior performance of the RCS in both systems.

An analysis of the average final scores by splitting methods shows an almost equal distribution between the two methods, with 148 entries for the TTS and 147 for the RCS, as indicated in **Table 3**. This balanced distribution provides a solid foundation for comparative analysis. The RCS consistently achieves higher average scores in both LMS and OAI retrieval methods, as shown in **Figure 3**. This result underscores the effectiveness of the RCS in maintaining contextual continuity within text chunks, contributing to its superior performance.

The ANOVA results, detailed in **Table 3**, reveal a statistically significant difference in retrieval performance between the two methods. For LMS scores, the p-value is less than 0.0001, and the same holds for OAI scores, indicating a substantial impact of the splitting method on retrieval effectiveness.

Given the observed advantages of the RCS in handling document structure and preserving context, further exploration and refinement of this method are recommended for broader applications in retrieval systems. Conversely, while the TTS proves useful, it may require adjustments to improve performance, particularly in maintaining contextual integrity.

**Table 3.** Comparative Analysis of Splitting Methods with Average Scores and ANOVA Results.

Splitting Method Analysis

| Splitting Method | Count | LMS Final Avg Score | OAI Final Avg Score |
|---|---|---|---|
| Recursive Character Splitter | 147 | 0.208 | 0.260 |
| Token Text Splitter | 148 | 0.192 | 0.223 |

ANOVA Results for LMS and OAI Scores by Splitting Method

| Source | Sum of Squares | df | F | p-value |
|---|---|---|---|---|
| C(Splitting Method) | 0.019 | 1 | 62.91572 | 4.587429e-14 |
| Residual | 0.088573 | 293 | | |
| C(Splitting Method) | 0.097836 | 1 | 46.976114 | 4.247097e-11 |
| Residual | 0.610223 | 293 | | |

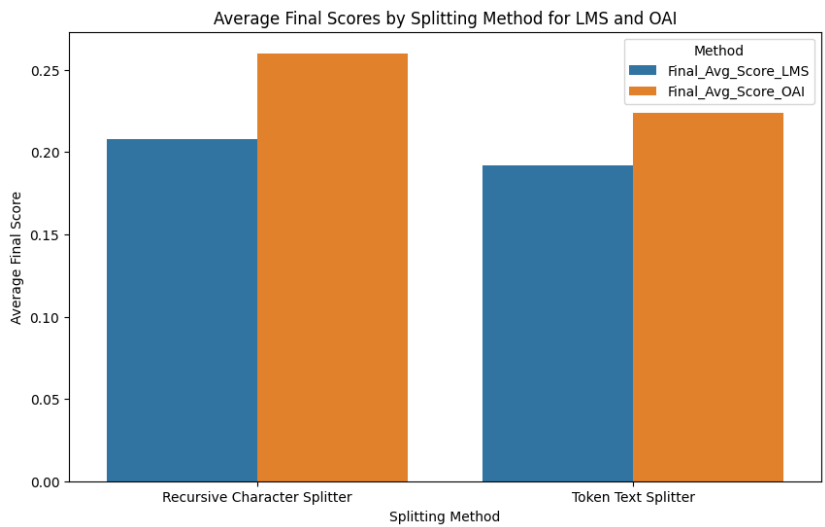

**Figure 3.** Average Final Scores by Splitting Method for LMS and OAI.

## 5.6 Analysis of Specific Cases with Pronounced Performance Differences

This analysis investigates specific cases with pronounced performance differences, as outlined in **Table 4**, highlighting the strengths and weaknesses of different document-splitting methods within the RAG system. In the top-performing cases using the RCS, both articles and textbooks achieved the highest differentials in LMS retrieval scores, demonstrating the effectiveness of RCS in managing complex document structures. Similarly, in OAI retrieval scores, textbooks and novels processed with RCS also exhibited high performance, underscoring the method's adaptability across different document types.

Conversely, the bottom-performing cases involving the TTS show that textbooks and novels recorded the lowest scores in LMS retrieval, reflecting the challenges TTS faces with complex or less structured content. This trend is mirrored in OAI retrieval scores, where textbooks and novels processed with TTS also displayed reduced performance, highlighting the limitations of this method in handling these document types.

The comparative analysis between RCS and TTS reveals that RCS superiorly supports retrieval outcomes by effectively preserving contextual integrity and structural coherence. RCS achieves this by adaptively recognizing and maintaining semantic structures within the text, ensuring that related concepts and thematic elements are not disrupted during segmentation. In contrast, TTS, which segments text primarily based on token counts, can inadvertently sever critical textual relationships, diminishing retrieval quality. These findings, supported by the performance metrics in **Table 4**, emphasize the importance of selecting a splitting method that aligns with the structural nuances of the documents and the specific objectives of the retrieval system, ensuring both relevance and coherence in the retrieved content.

**Table 4.** Detailed Performance Differences by Case and Method.

| Document_Type | Splitting_Method* | Final_Avg_Score_LMS | Score_Diff_LMS | Final_Avg_Score_OAI | Score_Diff_OAI |
|---|---|---|---|---|---|
| Article | RCS | 0.343926 | 0.136171 | - | - |
| Article | RCS | 0.321118 | 0.113363 | - | - |
| Textbook | RCS | 0.248542 | 0.052012 | - | - |
| Textbook | RCS | 0.244930 | 0.048400 | - | - |
| Article | RCS | 0.254614 | 0.046859 | - | - |
| Textbook | TTS | 0.161184 | -0.035346 | - | - |
| Novel | TTS | 0.166268 | -0.029770 | - | - |
| Novel | TTS | 0.167155 | -0.028883 | - | - |
| Novel | TTS | 0.168068 | -0.027970 | - | - |
| Textbook | TTS | 0.169725 | -0.026805 | - | - |
| Textbook | RCS | - | - | 0.469739 | 0.230228 |
| Textbook | RCS | - | - | 0.457366 | 0.217855 |
| Novel | RCS | - | - | 0.430353 | 0.204911 |
| Article | RCS | - | - | 0.456743 | 0.195836 |
| Novel | RCS | - | - | 0.399886 | 0.174443 |
| Novel | TTS | - | - | 0.168833 | -0.056610 |
| Textbook | TTS | - | - | 0.182959 | -0.056552 |
| Textbook | TTS | - | - | 0.183128 | -0.056383 |
| Novel | TTS | - | - | 0.169664 | -0.055779 |
| Novel | TTS | - | - | 0.169920 | -0.055522 |

*RCS (Recursive Character Splitter), TTS (Token Text Splitter)*

# 5.7 Performance-driven Content Structure Analysis

This analysis examines the cases from **Table 4** that exhibit significant performance differences, focusing on their content structures as detailed in **Table 5**. By correlating these content structures with the performance metrics from **Table 4**, the impact of varying structural elements on retrieval outcomes is elucidated. **Table 5** complements the findings from **Table 4** by providing a detailed breakdown of content features, offering insights into the factors contributing to both superior and suboptimal performance.

In the *top-performing cases* across both LMS and OAI retrieval systems, several content characteristics emerge as influential. Larger chunk sizes are consistently present, suggesting that more contextual information within larger chunks enhances retrieval performance by providing a comprehensive understanding of the content. Additionally, these cases exhibit a higher frequency of key terms, aligning with essential retrieval concepts and likely facilitating more accurate and relevant search results. Text complexity in these cases is maintained at a moderate level, striking a balance between readability and detail. This balance proves beneficial for effective information retrieval by providing sufficient depth without overwhelming the system with complexity.

Conversely, the *bottom-performing cases* demonstrate several limiting factors. These cases are characterized by smaller chunk sizes, which fail to provide adequate context for accurate data retrieval. Lower key term frequency is also a notable issue, reducing the system's ability to connect queries to relevant content, thereby diminishing retrieval performance. Text complexity in these cases tends to exhibit extremes—either too simplistic or overly complex. This range complicates the retrieval process, as overly simple content may lack critical information, while overly complex content can obscure key elements necessary for effective retrieval.

The findings from this analysis suggest that adaptive content strategies are necessary to optimize retrieval performance.

<u>Insights and Recommendations</u>

The variations observed highlight the need for adaptive content strategies that are specifically tailored to both the document type and retrieval objectives. By implementing enhancements such as optimizing chunk sizes, refining key term density, and adjusting text complexity, we can significantly enhance retrieval performance.

1. Optimize Chunk Generation: Implement dynamic adjustments to chunk sizes based on the specific characteristics of each document type to improve contextual coherence and retrieval accuracy.

2. Key Term Analysis: Conduct regular reviews of key term deployment strategies to ensure they are effectively facilitating the retrieval process.

3. Text Complexity Adjustments: Develop guidelines to balance text complexity, ensuring it is neither too simple to be trivial nor too complex to be impenetrable, thereby supporting effective and efficient information retrieval.

For a more detailed understanding of these observations, refer to **Table 5**, which presents a comparative analysis of the top- and bottom-performing cases across different document types, highlighting the impact of chunk size, key term frequency, and text complexity on retrieval outcomes.

**Table 5.** Detailed Content Characteristics of Top and Bottom Performing Cases.

| Document_Type | Chunk_Size | Key_Term_Frequency | Text_Complexity |
|---|---|---|---|
| Article | 358 | 19 | 6.039216 |
| Article | 298 | 16 | 5.644444 |
| Textbook | 171 | 8 | 5.880000 |
| Textbook | 135 | 6 | 4.913043 |
| Article | 599 | 29 | 5.315789 |
| Textbook | 211 | 10 | 6.066667 |
| Novel | 39 | 4 | 4.714286 |
| Novel | 42 | 2 | 5.142857 |
| Novel | 55 | 3 | 6.000000 |
| Textbook | 287 | 10 | 6.578947 |
| Textbook | 515 | 19 | 5.000000 |
| Textbook | 217 | 13 | 4.589744 |
| Novel | 351 | 11 | 5.175439 |
| Article | 590 | 34 | 4.737864 |
| Novel | 225 | 13 | 4.650000 |
| Novel | 55 | 3 | 6.000000 |
| Textbook | 195 | 8 | 4.939394 |
| Textbook | 213 | 18 | 4.631579 |
| Novel | 39 | 4 | 4.714286 |
| Novel | 44 | 4 | 5.428571 |

# 6. Conclusions

The EDA evaluated how different document types, textbooks, articles, and novels, affect the performance of the RAG system. Each document type presented unique challenges. Textbooks, with their structured and technical content, require specialized retrieval systems capable of handling complex layouts and terminologies. While textbooks are information-rich, their complexity can hinder retrieval if the system is not properly tuned. Articles, by contrast, are concise and thematically focused, often resulting in higher retrieval scores due to their straightforward structure, making them easier to manage for retrieval systems. Novels, characterized by their expansive narrative style, pose significant difficulties for retrieval, with fluid content and less defined structures often resulting in lower scores. These findings suggest that adaptive retrieval strategies, tailored to the characteristics of each document type, are crucial for optimizing accuracy and efficiency.

The analysis of document-splitting methods revealed substantial variations in retrieval performance. The Token-based Splitter, which segments documents based on tokens or patterns, is efficient for large volumes of text but struggles with complex documents requiring contextual continuity, such as novels and technical textbooks. In contrast, the Recursive Character Splitter (RCS) maintains the natural flow of the text, preserving contextual integrity and capturing nuanced information. This method consistently outperformed the Token-based Splitter across all document types, as confirmed by ANOVA results. The statistical significance of RCS's performance indicates its advantage in environments requiring precision and context preservation. These results emphasize the importance of selecting a splitting method aligned with both the document's structure and the system's retrieval goals.

The study compared two advanced retrieval methods, OpenAI and LM Studio, each demonstrating strengths in different contexts. OpenAI's retrieval, using the "text-embedding-3-small" model, achieved higher average scores, particularly excelling in articles where it demonstrated a 12% advantage over LM Studio. Its ability to capture semantic nuances contributed to its superior performance across all document types. LM Studio, employing the "bge-large-en-v1.5-gguf" model, excelled in handling structured documents like textbooks, performing well with technical terminology and complex structures. This differential performance highlights the sensitivity of retrieval systems to both document type and splitting method. A hybrid approach, combining the strengths of both systems, could enhance retrieval accuracy by leveraging OpenAI's semantic capabilities and LM Studio's proficiency in structured content retrieval.

Further analysis of content structure revealed key performance factors. Top-performing cases, across both LMS and OAI, featured larger chunk sizes, suggesting that providing more context within larger chunks improves retrieval performance. High key term frequency also contributed to better performance by aligning with critical retrieval concepts. Text complexity, while varied, tended to be moderate in top-performing cases, balancing readability with sufficient depth. In contrast, bottom-performing cases were characterized by smaller chunk sizes, resulting in a lack of context, and lower key term frequency, weakening the system's ability to match queries to content. Text complexity in these cases often fell at extremes (either too simple or overly complex) both of which detracted from effective retrieval.

Overall, this study underscores the need for adaptive retrieval strategies tailored to specific document characteristics, such as the structured nature of textbooks, the conciseness of articles, and the complexity of novels. The RCS consistently outperformed the Token-based Splitter, highlighting the importance of selecting a splitting method aligned with the document's structure. Further refinement of chunk and overlap sizes could enhance retrieval accuracy and efficiency. Future research should focus on optimizing these parameters across various document types. The introduction of a novel evaluation methodology, using an open-source model to generate a large dataset of question-and-answer pairs, simulates realistic retrieval scenarios and provides a robust framework for assessing RAG systems. This approach, combined with weighted scoring metrics (SequenceMatcher, BLEU, METEOR, and BERT Score), offers a comprehensive evaluation of system accuracy and relevance. Future work should apply and refine this technique across different document types and retrieval contexts to further improve RAG systems. By systematically testing and refining chunk configurations and retrieval strategies, this research provides a roadmap for optimizing RAG systems for modern data challenges, contributing to advancements in information retrieval. The code used for this research is open-source and can be accessed at GitHub repository.

# Acknowledgments

The authors would like to express their gratitude to Professor Eduard Hovy, Executive Director at Melbourne Connect, University of Melbourne, for his insightful advice and guidance, which significantly contributed to the development of this work.